\documentclass[11pt]{article}
\def\ket#1{|{#1}\rangle}

\def\Qop#1{
\begin{picture}(4,1.5)(0,0.5)
\put(0,0.75){\line(1,0){1}}
\put(1,-.25){\framebox(2,2){$#1 $}}
\put(3,0.75){\line(1,0){1}}
\end{picture}}

\def\Qpass{
\begin{picture}(4,1.5)(0,0.5)
\put(0,0.75){\line(1,0){4}}
\end{picture}}

\begin{document}
\title{Quantum Computing}
\author{Eleanor Rieffel \\ FX Palo Alto Laboratory}
\date{}
\maketitle

\tableofcontents


\begin{abstract}
Changing the model underlying information and computation from a
classical mechanical to a quantum mechanical one yields faster 
algorithms, novel cryptographic mechanisms, and alternative methods 
of communication. Quantum algorithms can perform a select set of tasks
vastly more efficiently than any classical algorithm, but for many
tasks it has been proven that quantum algorithms provide no advantage.
The breadth of quantum computing applications is still being explored.
Major application areas include security and the many fields that 
would benefit from efficient quantum simulation. The quantum information
processing viewpoint provides insight into classical algorithmic
issues as well as a deeper understanding of entanglement and other
non-classical aspects of quantum physics.
\end{abstract}

\pagebreak

\section{Introduction}

Quantum computation explores how efficiently nature allows us
to compute. The standard model of computation is grounded
in classical mechanics; the Turing machine is described in 
classical mechanical terms. In the last two decades of the 
twentieth century, researchers recognized that the standard 
model of computation placed unnecessary limits on computation. 
Our world is inherently quantum mechanical. By placing 
computation on a quantum mechanical
foundation faster algorithms, novel cryptographic mechanisms,
and alternative methods of communication have been found.
Quantum information processing, a field that includes quantum
computing, quantum cryptography, quantum communication, and
quantum games, examines the implications of using a quantum mechanical
model for information and its processing.
Quantum information processing changes not only the physical processes
used for computation and communication, but the very notions of information 
and computation themselves.

Quantum computing is not synonymous with using quantum effects
to perform computation. Quantum mechanics has been an integral
part of modern classical computers and communication devices 
from their earliest days, the transistor and the laser being the
most obvious examples.
The phrase ``quantum computing'' is not parallel with
the phrases ``DNA computing'' or ``optical computing'': these describe 
the substrate on which computation is done without changing the
notion of computation. The phrase ``quantum computing'' is closer
in character to ``analog computing'' because the computational model
for analog computing differs from that of standard computing: a continuum
of values is allowed, rather than only a discrete set. While 
the phrases are parallel, the two models differ greatly.
The fundamental unit of quantum computation, the qubit, 
can take on a continuum of values, but a discrete version of quantum
computation can be constructed that preserves the features of standard
quantum computation.

Quantum computing does not provide efficient solutions to 
all problems. Nor does it provide
a universal way of circumventing the slowing of Moore's law as 
fundamental limits to miniaturization are reached.
Quantum computation enables certain
problems to be solved efficiently; some problems which on
a classical computer would take more than the age of the universe, 
a quantum computer could solve in a couple of days. But for
other problems it has been proven that quantum computation cannot
improve on classical methods, and for yet another class, that the
improvement is small. Quantum computation
will have significant impact on security and the
many fields which will benefit from faster and more accurate quantum
simulators.

\section{Early history}

In the early 1980s, Feynman, Manin, and others
recognized that certain quantum phenomena - phenomena associated with
entangled particles - could not be simulated efficiently
on standard computers. Turning this observation around, researchers
wondered whether these quantum phenomena could be used to speed up computation
in general. Over the next decade, a small group of researchers
undertook the task of rethinking the models underlying information 
and computation and providing formal models.
Deutsch developed a notion of a quantum mechanical Turing machine.
Bernstein, Vazirani, and Yao showed that quantum computers
can do anything a classical computer can do with at most
a small (logarithmic) slow down.

The early 1990s saw the first
truly quantum algorithms, algorithms with no classical analog 
that were provably better than any possible classical algorithm. 
The first of these was Deutsch's algorithm, later generalized to the
Deutsch-Jozsa algorithm. 
These initial quantum algorithms were able to
solve problems efficiently with certainty 
that classical techniques can solve efficiently only with  
high probability. Such a result is of no practical interest since 
any machine has imperfections so can only solve problems
with high probability. Furthermore, the problems solved were
highly artificial. Nevertheless,
such results were of high theoretical interest since they proved 
that quantum computation is theoretically more 
powerful than classical computation. 

These results inspired Peter Shor's 1994 polynomial-time
quantum algorithm for factoring integers. This result provided a 
solution to a well-studied problem of practical interest.
A classical polynomial-time solution has long eluded researchers.
Many security protocols 
base their security entirely on the computational
difficulty of this problem. 
Shor's factoring algorithm and related results mean that
once a large enough quantum computer is built, all standard public
key encryption algorithms will be completely insecure. 

Shor's results sparked interest in the field, but
doubts as to its practical significance remained. Quantum
systems are notoriously fragile. Key quantum properties, such as 
entanglement, are easily disturbed by environmental
influences. 
Properties of quantum mechanics, such as the impossibility of reliably
copying an unknown quantum state, made it look unlikely that 
error correction techniques for quantum computation could
ever be found. 
For these reasons, it seemed unlikely that reliable quantum 
computers could be built.
Luckily, in spite of widespread doubts as to whether
quantum information processing could ever be made practical, the theory
itself proved so tantalizing that researchers continued to explore it.
In 1996 Shor and Calderbank, and independently Steane, developed
quantum error correction techniques.
Entanglement provides a key resource.
Today, quantum error correction is arguably the most mature
area of quantum information processing. 

The notions underlying quantum computation are highly technical and not
easily explained because they rely on unintuitive aspects of quantum
mechanics that have no classical analog. The next section briefly
introduces a few of the most fundamental concepts. The following
sections discuss the applications of quantum computation, its
limitations, and the efforts to build quantum information processing
devices.

\section{Basic concepts of quantum computation}

The state space of a physical system consists of all possible states 
of the system. Any quantum mechanical system that can be modeled by 
a two dimensional complex vector space can be viewed as a qubit.
Such systems include photon polarization, electron spin, and a 
ground state and an excited state of an atom. 
A key difference between classical and quantum systems is the
way in which component systems combine. 
The state of a classical system can be
completely characterized by the state of each of
its component pieces.
A surprising and unintuitive aspect of quantum systems is
that most states cannot be
described in terms of the states of the system's components.
Such states are called {\it entangled states}. 

Another key property is {\it quantum measurement}.
In spite of there being a continuum of possible states, any
measurement of a system of qubits has only a discrete set
of possible outcomes; for $n$ qubits, there are at most $2^n$
possible outcomes. After measurement, 
the system will be in one of the possible outcome states.
Which outcome is obtained is probabilistic; outcomes closest to 
the measured state are most probable. 
Unless the state is already in one of the possible
outcome states, measurement changes
the state; it is not possible to reliably measure an unknown state
without disturbing it. 

Just as each measurement has a discrete set of possible outcomes, 
any mechanism for copying quantum states can only correctly copy
a discrete set of quantum states. For an $n$ qubit system, 
the largest number of quantum states a copying mechanism can copy
correctly is $2^n$.
For any state there is a mechanism that can correctly copy it, but
if the state is unknown, there is no way to determine which mechanism
should be used. For this reason, it is impossible to copy reliably an unknown
state, an aspect of quantum mechanics called the 
{\it no cloning principle}. 

A qubit has two arbitrarily chosen distinguished states,
labeled $\ket 0$ and $\ket 1$, 
which are the possible outcomes of a single measurement.
Every single qubit state can be represented as a linear combination,
or {\it superposition}, of these two states.
In quantum information processing, classical bit values of $0$ and $1$
are encoded in the distinguished states $\ket 0$ and $\ket 1$.
This encoding enables a direct comparison between bits and qubits: bits
can only take on two values, $0$ and $1$, while qubits can take on 
any superposition of these values, $a\ket0 + b\ket 1$, where $a$ 
and $b$ are complex numbers such that $\vert a\vert^2 + \vert b\vert^2 = 1$.

Any transformation of an $n$ qubit system can be obtained
by performing a sequence of one and two qubit operations. Most 
transformations cannot be performed efficiently in this manner. 
Figuring out an efficient sequence of quantum transformations
that can solve a useful problem is the heart of quantum
algorithm design.

\section{Quantum algorithms}

Problems generally get harder as the size of the input increases.
The efficiency of an algorithm is quantified in terms of an
asymptotic quantity that looks at how the resources used by the
algorithm grow with the input. Time and space, 
generally measured in terms of number of operations
and number of bits or qubits, are the resources most often
considered. Constant factors are usually ignored, since they depend 
on fine details of an implementation that often are not known, but 
can be bounded. An algorithm is polynomial in the input size $n$ if 
the amount of resources used is less than a fixed polynomial of $n$; 
in such a case the algorithm is said to be $O(n^k)$ for some $k$, 
the degree of a bounding polynomial. Algorithms whose
resource use cannot be bounded by a polynomial are said to be
superpolynomial. Algorithms whose resource use is 
asymptotically greater than some exponential function of $n$ are 
said to be exponential.  Algorithms of the same polynomial degree
are generally viewed as achieving the same level of efficiency.

It is easy to take a {\it reversible} classical computation 
and turn it into an equivalently efficient quantum 
computation. Bennett showed in 1973 that any classical computation 
using $t$ time and $s$ space has a reversible
counterpart using only $O(t^{1+\epsilon})$ time and
$O(s\log t)$ space. Thus for every classical computation there
is a quantum computation of similar efficiency. 
Truly quantum algorithms use other methods to solve
problems more efficiently than is possible classically.
Discovering novel approaches remains an active but difficult area of 
research. After 1996, there was a hiatus of five years
before a significantly new algorithm was discovered. Then alternative
models of quantum computation and quantum random walks inspired
new types of algorithms.

Most researchers expect that quantum computers cannot solve $NP$-complete
problems in polynomial time. Informally, a problem is in $NP$ if there is
an efficient way to check that a proposed solution is a solution. 
A problem is in $P$ if a solution can be found in polynomial time. 
A problem is $NP$-complete if an efficient solution to that problem 
would imply an efficient solution to all problems in $NP$.
There is no proof that quantum computers cannot solve $NP$-complete
problems in polynomial time (a proof would imply $P\ne NP$, a long standing
open problem in computer science). Ladner's theorem says that
if $P\ne NP$, then there exist $NP$ intermediate problems:
problems that are in $NP$, and not in $P$, but are not $NP$ complete.
Candidate $NP$ intermediate problems include factoring and the discrete
logarithm problem.  Other candidate problems include graph isomorphism,
the gap shortest lattice vector problem, and many hidden subgroup
problems. Whether there are polynomial quantum algorithms for 
these other problems remains a major open question.

\subsection{Grover's algorithm and generalizations}

Grover's search algorithm is the most famous quantum algorithm after
Shor's algorithm. It searches an unstructured list of $N$ items
 in $O(\sqrt N)$ time. The best possible classical algorithm
uses $O(N)$ time. This speed-up is small but, unlike for Shor's
algorithm, it has been proven that Grover's algorithm outperforms any possible
classical approach.  Although Grover's original algorithm succeeds only
with high probability, variations that succeed with certainty are known;
Grover's algorithm is not inherently probabilistic.

Generalizations of Grover's algorithm apply
to a more restricted class of problems than is generally realized.
It is unfortunate that Grover used ``database" in the title of his 
1997 paper. Databases are generally highly structured and can be 
searched rapidly classically. Because Grover's algorithm does not 
take advantage of structure in the data, it does not provide
a square root speed up for database search.
Childs et al. showed that quantum computation 
can give at most a constant factor improvement for searches of 
ordered data like that of databases.
Furthermore, Grover's algorithm destroys the
quantum superposition of the data, so the
superposition must be recreated for each search. This recreation
is often linear in $N$
which negates the $O(\sqrt N)$ benefit of Grover's algorithm, reducing
its applications still further; 
the speed-up is obtained only for data that has a sufficiently fast
generating function. 

Extensions of Grover's algorithm 
provide small speed-ups for a variety of problems
including approximating the mean of a sequence and other statistics,
finding collisions in $r$-to-$1$ functions, string matching, 
and path integration.  Grover's algorithm has also
been generalized to arbitrary initial conditions, 
non-binary labelings, and nested searches.

\subsection{Generalizations of Shor's factoring algorithm}

At the same time Shor discovered his factoring algorithm, he also
found a polynomial time solution for the discrete logarithm 
problem, a problem related to factoring that is also
heavily used in cryptography. Both factoring and the discrete 
logarithm problem are {\it hidden subgroup problems}. In particular,
they are both examples of abelian hidden subgroup problems. Shor's
techniques are easily extended to all abelian hidden subgroup
problems and a variety of hidden subgroup problems over 
groups that are close to being abelian.

Two cases of the hidden subgroup problem have received a lot
of attention: the symmetric group $S_n$, the full permutation group
of $n$ elements, and the dihedral group $D_n$, the group of symmetries
of a regular $n$-sided polygon. But efficient algorithms have eluded
researchers so far.
A solution to the hidden subgroup problem over $S_n$ would yield 
a solution to graph isomorphism, a prominent NP intermediate
candidate. In 2002, Regev showed that an efficient algorithm to the 
dihedral hidden subgroup problem using Fourier sampling, a generalization
of Shor's techniques, 
would yield an efficient algorithm for the gap
shortest vector problem. Public key cryptographic schemes based on
shortest vector problems are among the most promising approaches
to finding practical public key cryptographic schemes that are 
secure against quantum computers.  In 2003, Kuperberg found a
subexponential (but still superpolynomial) algorithm for the dihedral group.

Efficient algorithms have been obtained for some related problems.
In 2002, Hallgren found an efficient quantum algorithm for solving 
Pell's equation. Pell's equation, believed to be harder than 
factoring and the discrete logarithm problem, was the security basis for 
Buchmann-Williams key exchange and public key cryptosystems.
Thus Buchmann-Williams joins the many public key cryptosystems 
known to be insecure in a world with quantum computers.
In 2003, van Dam, Hallgren, and Ip found an efficient quantum algorithm for
the shifted Legendre symbol problem, which means
that quantum computers can break certain algebraically homomorphic
cryptosystems and can predict certain pseudo-random number generators.

\subsection{Other classes of algorithms}

In 2002, a new family of quantum algorithms
emerged that uses quantum random walk techniques
to solve a variety of problems
related to graphs, matrix products, and relations in groups. 
The alternative models of quantum computation that will be 
discussed in section \ref{models}, such as cluster state and adiabatic
quantum computing, led to other novel types of
quantum algorithms.

\subsection{Simulation}

Simulation of quantum systems is another major application of quantum 
computing; it was the recognition of the difficulty
of simulating certain quantum systems that started the field of
quantum computation in the first place.
Already, in the early 2000s, small scale quantum simulations
have provided useful results.
Simulations run on special purpose quantum devices will have
applications in fields ranging from chemistry, to biology, to
material science. They will also support the design and implementation
of yet larger special purpose quantum devices, a process that ideally
leads all the way to the building of scalable general purpose
quantum computers.

Even on a universal 
quantum computer, there are limits to what information can
be gained from a simulation. Some quantities, like the energy spectra
of certain systems, are exponential in quantity, so no algorithm, 
classical or quantum, can output them efficiently. 
For other quantities, algorithmic advances are needed
to determine whether and how  that information can be efficiently
extracted from a simulation. 

Many quantum systems can be efficiently simulated classically. After all,
we live in a quantum world but have long been able to 
simulate a wide variety of natural 
phenomena. Some entangled quantum systems can be efficiently simulated
classically, while others cannot. The question of exactly which 
quantum systems can be efficiently
simulated classically remains open. New approaches to classical
simulation of quantum systems continue to be developed, many benefiting
from the quantum information processing viewpoint.
The quantum information processing viewpoint has also led to improvements
in a commonly used classical approach to simulating quantum systems,
the density matrix renomalization (DMRG) approach.

\section{Limitations of quantum computing}

Beals et al. proved that, for a broad class of problems,
quantum computation cannot provide any speed-up. Their methods
were used by others to provide lower bounds for other types of
problems. Ambainis found another powerful
method for establishing lower bounds.
In 2002, Aaronson showed that quantum approaches could not be used
to efficiently solve collision problems.
This result means there is no generic quantum attack
on cryptographic hash functions. 
Shor's algorithms break some cryptographic hash functions,
and quantum attacks on others may still be discovered, but Aaronson's
result says that any attack must use specific properties of the
hash function under consideration. 

Grover's search algorithm is optimal; it is not possible to
search an unstructured list of $N$ elements more rapidly than 
$O(\sqrt{N})$. This bound was known before Grover found his algorithm.
Childs et al. showed that for ordered data,
quantum computation can give no more that a constant factor improvement
over optimal classical algorithms.
Grigni et al. showed in 2001 that for
most non-abelian groups and their subgroups, the standard Fourier sampling
method, used by Shor and successors, yields exponentially little
information about a hidden subgroup. 

\section{Quantum protocols}

Applications of quantum information processing include a number
of communication and cryptographic protocols. The two most famous
communication protocols are quantum teleportation and dense coding.
Both use entanglement shared between the two parties that are
communicating.  Teleportation uses two classical bits, together
with shared entanglement, to transmit the state of a single 
qubit. It is surprising that
two classical bits suffice to communicate 
any one of an infinite number of possible single qubit states.
Teleportation destroys the state at the original
site in the process, leading to the name teleportation. 
In this way, teleportation enables the transmission of an unknown 
quantum state without violating the no-cloning principle.
Dense coding uses one quantum
bit, together with shared entanglement, to transmit two
classical bits.  Since the entangled particles can be distributed 
ahead of time, only one qubit needs to be physically transmitted 
to communicate two bits of information. This result is surprising since
only one classical bit's worth of
information can be extracted from a qubit. Both protocols show
the power of a small amount of shared entanglement.

Quantum key distribution schemes were the first examples of quantum 
protocols. The first scheme, due to Bennett and Brassard in 1984, 
uses properties of quantum measurement; no entanglement is needed. 
Quantum key distribution protocols perform the same function
as the classical Diffie-Hellman key agreement protocol, to establish
a secret symmetric key between both parties, but
their security rests on properties of
quantum mechanics. The Diffie-Hellman protocol relies on the
computational intractability of the discrete logarithm problem;
it remains secure against all known classical attacks, but 
is broken by quantum computers. Other quantum key distributions schemes
exist, including Ekert's entanglement based scheme. Many of the
schemes have been demonstrated experimentally, over fiber
optic cable and in free space.
Three companies, id Quantique, MagiQ, and SmartQuantum,
focus on quantum cryptography, while a number of other companies, including
BBN, NTT, NEC, Mitsubishi, and Toshiba, have contributed to the area. 

While ``quantum cryptography'' is often used as a synonym for
``quantum key distribution,'' quantum approaches to a wide variety of other
cryptographic tasks have
been developed.  Some of these protocols 
use quantum means to secure
classical information. Others secure quantum information. Many
are ``unconditionally" secure in that their security is based 
entirely on properties of quantum mechanics. Others are only
quantum computationally secure in that their security depends on 
a problem being computationally intractable for a quantum
computers. For example, while ``unconditionally" secure bit
commitment is known to be impossible to achieve through either
classical or quantum means, quantum computationally
secure bit commitments schemes exist as long as there are  
quantum one-way functions.

Closely related to quantum key distribution schemes are protocols for 
unclonable encryption, a symmetric
key encryption scheme that guarantees that an eavesdropper cannot
copy an encrypted message without being detected. Unclonable
encryption has strong ties with quantum authentication. 
One type of authentication is digital signatures. 
Quantum digital signature schemes have been developed,
but the keys can be used only a limited number of times. 
In this respect they resemble classical schemes such as 
Merkle's one-time signature scheme. 

Cleve et al. provide quantum protocols
for $(k,n)$ threshold quantum secrets. Gottesman et al. 
provide protocols for more general quantum secret sharing. 
Quantum multiparty function evaluation schemes
exist. 
Fingerprinting enables the 
equality of two strings to be determined efficiently with high probability by 
comparing their respective fingerprints. 
Classical fingerprints for $n$ bit strings need to be at 
least of length $O(\sqrt{n})$.  Buhrman et al. show 
that a quantum fingerprint of classical data can be exponentially 
smaller.

In 2005, Watrous showed that many classical
zero knowledge interactive protocols are zero knowledge against a 
quantum adversary. 
Generally, statistical zero knowledge protocols
are based on candidate NP-intermediate problems,
another reason why zero knowledge protocols
are of interest for quantum computation.
There is a close connection between quantum interactive protocols
and quantum games. Early work by Eisert et al. includes a discussion of
a quantum version of the prisoner's dilemma. 
Meyer has written lively papers discussing other quantum games.

\section{Broader implications of quantum information processing}

Quantum information theory has led to insights into fundamental
aspects of quantum mechanics, particularly entanglement. 
Efforts to build quantum information processing devices have resulted 
in the creation of highly entangled states that have enabled deeper
experimental exploration of quantum mechanics. These entangled states, and 
the improvements in quantum control, have been used in quantum 
microlithography to 
affect matter at scales below the wavelength limit and in quantum
metrology to achieve extremely accurate sensors. Applications include
clock accuracy beyond that of current atomic clocks, which are limited 
by the quantum noise of atoms, optical resolution beyond the wavelength 
limit, ultra-high resolution spectroscopy, and ultra-weak absorption 
spectroscopy.

The quantum information processing viewpoint has also provided a new
way of viewing complexity issues in classical computer science, and
has yielded novel classical algorithmic results and methods.
Classical algorithmic results stemming from the insights of quantum
information processing include lower bounds for problems involving 
locally decodable codes, local search, lattices, 
reversible circuits, and matrix rigidity. 
The usefulness of the complex perspective for evaluating
real valued integrals is often used as an analogy to explain
this phenomenon. We examine one example of an application of
quantum information processing to classical
computer science.

Cryptographic protocols usually rely on the empirical hardness of
a problem for their security; it is rare to be able to prove
complete, information theoretic security. When a cryptographic protocol
is designed based on a new problem, the
difficulty of the problem must be established before the security
of the protocol can be understood. Empirical testing of a problem
takes a long time. Instead, whenever possible,
``reduction" proofs are given that show that if the new problem were
solved it would imply a solution to a known hard problem.
Regev designed a novel, purely classical cryptographic
system based on a certain lattice problem. He was able to reduce 
a known hard problem to this problem, but only by using a quantum step
as part of the reduction proof.

\section{Impact of quantum computers on security}
\label{PKEandQC}

Electronic commerce relies on secure public key encryption and digital 
signature schemes, as does secure electronic communication.  
Public key encryption is used to authenticate the communicating parties,
and to distribute symmetric session keys, the
keys used to encode data for transmission.
Public-private key pairs consist of a public key, knowable by all
and therefore easy to distribute,
and a corresponding private key whose secrecy must be maintained.
Symmetric keys consist of a single key
(or a pair of keys easily computable from one another) that are known
only to the legitimate parties. 
Without secure public key encryption, authentication and the distribution 
of symmetric session keys become unwieldy.

Public key encryption is the digital equivalent of a locked mailbox:
anyone can put a message in, but only the person with the key can
read the message. Public key encryption schemes have
digital analogs of the locked box and the key.
Publicly known {\it one way functions} provide the digital analog of a 
locked box: they are easy to compute, but the inverse function is
hard to compute, just as it is easy to put a letter 
in a locked mailbox, but hard to get it out again without the
key. The digital analog of the key is a {\it trapdoor}, additional
information that makes the inverse easy to compute.

All practical public key encryption protocols use one-way trapdoor
functions based on either factoring 
or the discrete logarithm problem. RSA, Rabin, 
Goldwasser-Micali, LUC, Fiege-Fiat Shamir, ESIGN, SSL, https rely
on factoring, while Diffie-Hellmen, DSA, El Gamal, and elliptic curve
cryptography rely on the discrete logarithm problem. 
Shor's quantum algorithms render all of these encryption schemes insecure
by providing a means of computing the inverse function almost as easily
as the original function. Once quantum computers have been built,
what were one-way trapdoor functions are no longer one-way.
Limited use classical or quantum signature schemes, such as Merkle's or
Gottesman's, provide only an inefficient substitute. 
So if scalable 
quantum computers existed today, the world would not have a secure 
means for efficient electronic commerce. 

Even before Shor discovered his algorithms, cryptographers were worried
about the dependence of public key encryption on just two closely related 
problems. However, developing
alternative public key algorithms based on other mechanisms has
proven difficult.  
McEliece is not practical; for the recommended security parameters 
the public key size is $2^{19}$ bits, and because of its impracticality, 
its security has received less scrutiny than had the protocol been
more practical. 
All knapsack-based public key cryptosystems have been broken, including 
the Chor-Rivest scheme which stood for 13 years. Many other types of 
public key cryptosystems have been developed and then broken.

Both factoring and the discrete logarithm problem are candidate
NP intermediate problems. Hope for alternative
public key encryption protocols centers on using other NP 
intermediate problems. The leading candidates 
are certain lattice based problems. 
Some of these schemes have impractically large keys, while for
others their security remains in question. Also, Regev showed 
that lattice based problems are closely related to the dihedral
hidden subgroup problem. The close relationship of the dihedral
hidden subgroup problem with problems solved by Shor's algorithm 
makes many people nervous, though so far the dihedral hidden
subgroup problem has resisted attack.

Given the historic difficulty of creating practical public key
encryption systems
based on problems other than factoring or discrete log, it is unclear
which will come first, a large scale quantum computer or
a practical public key encryption system secure against
quantum and classical attacks. If the building of quantum computers
wins the race, the security of electronic commerce and communication
around the world will be compromised.

\section{Implementation efforts}

DiVincenzo developed widely used requirements for a quantum computer.
It is relatively easy to obtain $N$ qubits, but 
it is hard to get them to interact with each other and 
with control devices, but nothing else.
DiVincenzo's criteria are, roughly:
\begin{itemize}
\item Scalable physical system with well-characterized qubits
\item Ability to initialize the qubits in a simple state
\item Robustness to environmental noise
\item A set of ``universal" gates that approximate all quantum operations 
\item High efficiency, qubit-specific measurements 
\end{itemize}

There are daunting technical difficulties in actually building such a machine.  
Research teams around the world are actively studying ways to build practical 
quantum computers.  The field is changing rapidly. It is impossible
even for experts to predict which of the many approaches are likely
to succeed. As of 2008, no one has made a detailed proposal that meets 
all of the DiVincenzo criteria, let alone realize it in a laboratory. 
Many promising approaches are being pursued by theorists and experimentalists
around the world. Researchers are actively exploring various architectural 
needs of and designs for quantum computers and evaluating
different quantum technologies for achieving these needs.
A breakthrough will be needed to go beyond tens of
qubits to a quantum computer meeting DiVincenzo's criteria 
with hundreds of qubits. 

The earliest small quantum computers used liquid nuclear magnetic
resonance (NMR) technology that was already highly advanced due to 
its use in medicine. A quantum bit is encoded in the average spin state of 
a large number of nuclei of a molecule. Each qubit corresponds to a particular  
atom of the molecule; the qubits can be distinguished from each
other by the nucleus of their atom's characteristic frequency.  
The spin states can be manipulated by magnetic fields and the average 
spin state can be measured with NMR techniques.
Liquid NMR appears unlikely to lead implementation
efforts much longer, let alone achieve a scalable quantum computer,
due to severe scaling problems; the measured signal drops off
exponentially with the number of qubits.

The history of optical approaches to building a quantum computer
illustrates how hard it is to make good predictions.
Optical methods are the unrivaled approach for quantum communications
applications because photons do not interact with much. 
This same trait, however, means
that it is difficult to get photons to interact with each other,
which made them appear unsuitable as the fundamental qubits on
which computation would be done. So in 2000 optical approaches
were considered unpromising.
While ``nonlinear" optical materials induce some photon-photon 
interactions, no known material has a sufficiently strong non-linearity,
and scientists doubt such a material will ever be found.
In 2001, Knill, Laflamme and Milburn (KLM) showed how, by clever 
use of measurement, non-linear optical elements could be avoided
altogether. However, the overhead
was enormous. In 2004, Nielsen reduced this overhead
by combining the KLM approach with cluster state quantum computing.

In an ion-trap quantum computer individual ions,
confined by electric fields, represent single qubits.
Lasers directed at ions perform single qubit operations 
and two qubit operations between adjacent ions.  All operations
necessary for quantum computation have been demonstrated in the laboratory for
small numbers of ions. To scale this technology, proposed architectures
include quantum memory and processing elements where qubits are moved back
and forth either through physical movement of the ions
or by using photons to transfer their state. 
Many other approaches exist, including cavity QED, neutral atom,
Josephson junctions, and and various other solid state 
approaches. Hybrid approaches are also
being pursued. Of particular interest are interfaces between optical
qubits and other forms.

Once a quantum information processing device is built, it must be
tested to see if it works as expected and to determine
what sorts of errors occur. Finding efficient
methods of testing is a challenge, given the
large state space and the effects of measurement on the system.
{\it Quantum state tomography}
provides procedures for experimentally characterizing a quantum state.
{\it Quantum process tomography} experimentally characterizes a sequence
of operations performed by a device. 

\section{Advanced concepts}

\subsection{Robustness}

Environmental interactions muddle quantum computations.
It is difficult to isolate a quantum computer sufficiently
from environmental interactions, especially because
controlled interactions are needed to perform the computation.
In the classical world, error correcting codes are primarily used in
data transmission. But the effects of the environment on
any quantum information processing device 
are likely to be so pervasive that quantum states will need protection
at all times. 

Fault tolerant techniques limit the propagation of errors during 
computation to keep them manageable enough that quantum error correction
techniques can handle them. Fault tolerant error correction techniques
make sure that even if the error correction process is faulty, it 
introduces fewer errors than it cures. Powerful threshold
theorems have been proven; a quantum computer with an
error rate below a certain threshold can run arbitrarily
long computations to whatever accuracy is desired.
Threshold results exist for a variety of codes and error 
models.

Alternative approaches to robust quantum computation exist. 
Instead of encoding so that common errors can be detected and corrected,
all computation can be performed in subspaces unaffected
by common errors. These
``decoherence-free subspace'' approaches are
complementary to error correcting codes. 
Operator error correction provides a
framework that unifies quantum error correcting codes and
decoherence-free subspaces.
Quantum computers built according to the topological model of quantum
computation have innate robustness. 
Most likely, actual quantum computers will use quantum error correcting 
codes in combination with other approaches.

\subsection{Models underlying quantum computation}
\label{models}

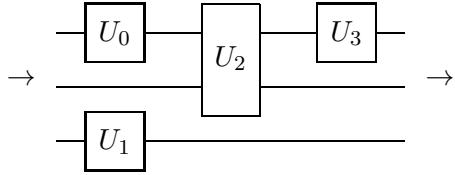
\begin{figure}
\[\to
\begin{array}{c}
\Qop{U_0} 
        \begin{picture}(4,1.5)(0,0.5)
        \put(0,0.75){\line(1,0){1}}
        \put(1,1.75){\line(2,0){2}}
        \put(1,1.75){\line(0,-1){3}}
        \put(3,1.75){\line(0,-1){3}}
        \put(3,0.75){\line(1,0){1}}
        \end{picture}
          \Qop{U_3}\\
\Qpass  
        \begin{picture}(4,1.5)(0,0.5)
        \put(0,0.75){\line(1,0){1}}
        \put(1,-.25){\line(2,0){2}}
        \put(1,-.25){\line(0,1){3}}
        \put(2, 1.75){\makebox(0,0){$U_2$}}
        \put(3,-.25){\line(0,1){3}}
        \put(3,0.75){\line(1,0){1}}
        \end{picture}
   \Qpass   \\
\Qop{U_1} \Qpass  \Qpass   \\
  \end{array} \to\]
\caption{A graphical representation for a $3$-qubit quantum circuit.
Each horizontal line represents a qubit. Time runs from left to right.
The boxes represent basic one and two qubit quantum gates applied to
the appropriate qubits.}
\label{Qcircuit}
\end{figure}

A {\it circuit model} for universal quantum computation consists of 
a set of one and two qubit transformations, {\it quantum gates},
from which all quantum transformation can be approximated. 
Circuit diagrams such as the one shown in figure \ref{Qcircuit}
are often drawn, but these should not be taken literally; these
are not blueprints for quantum hardware, but rather abstract
diagrams indicating a sequence of operations to be performed. 
Each horizontal line represents a qubit. Time runs from left to 
right, and the boxes represent one and two qubit quantum gates applied
to the qubits. In an ion-trap quantum computer, 
these diagrams indicate the sequence of laser pulses to apply.
Because efficiency of a quantum algorithm can be quantified in 
terms of the number of qubits and basic transformations used, 
and because there are quantum gates corresponding
to basic classical logic operations, this model enables
a direct comparison of quantum and classical algorithms, and makes 
finding quantum analogs of classical computation straightforward.

It is less clear that the circuit model is ideal for inspiring
new quantum algorithms or giving insight into 
the limitations of quantum computation. Other models may give
more insight into quantum algorithmic design or the physical
realization of quantum computers and their robustness. 
Two alternative models of quantum computation have proven 
particularly fruitful: cluster state quantum computation and adiabatic 
quantum computation.

Cluster state quantum computation illuminates the entanglement 
resources needed for quantum computation.
In cluster state, or one-way, quantum computing
a highly entangled ``cluster" state is set up at the beginning
of the algorithm. All computations take place by single qubit
measurements, so the entanglement between the qubits can only decrease
in the course of the algorithm (the reason for the ``one-way" name).
The initial cluster state is independent of the
algorithm to be performed; it depends only on the size of the problem
to be solved. In this way cluster state quantum computation makes
a clean separation between the entanglement creation and 
computational stages. 
Cluster state quantum computing underlies the most promising approaches
to optical quantum computation. 

Adiabatic quantum computation rests on the Hamiltonian framework
for quantum mechanics. A problem is encoded in the Hamiltonian
of a system in such a way that a solution is a ground state. 
An adiabatic algorithm begins with the system in the ground state of 
an easily implementable Hamiltonian. The Hamiltonian is gradually 
perturbed along a path 
between the initial Hamiltonian and the problem Hamiltonian.
The adiabatic theorem says that if
the path is traversed slowly enough the system will remain in a
ground state, so at the end of computation it will be in a 
solution state.  How slowly the path must be traversed depends on 
spectral properties of the Hamiltonians along the path.
Which Hamiltonians can be used affects the computational power. Some
versions of adiabatic computation are equivalent
to quantum computation, but others are only
classical. Small adiabatic computational
devices have been built; in some cases it has not been possible
to determine whether they perform quantum computation or not.
Initial interest centered on the possibility of using adiabatic
computation to solve NP-complete problems,
because adiabatic algorithms were not subject to the lower bound results 
proven for other approaches. Vazirani and van Dam, and later
Reichardt, were able to rule out a variety of such adiabatic approaches.  
Quantum adiabatic solutions to other problems have been found.

Holonomic, or geometric, quantum computation is a hybrid between 
adiabatic quantum computation and the circuit model in which the 
quantum gates are implemented via adiabatic processes. Holonomic
quantum computation makes use of non-Abelian geometric phases that 
arise from perturbing a Hamiltonian adiabatically along a loop in 
its parameter space. The phases depend only on topological properties 
of the loop so are insensitive to perturbations. This property means that 
holonomic quantum computation has good robustness with respect
to errors in the control driving the Hamiltonian's evolution.
Early experimental efforts have been carried out using a variety of
underlying hardware.

In 1997, prior to the development of the holonomic approach to
quantum computing, Kitaev proposed topological quantum computing,
a more speculative approach to quantum computing with great robustness
properties. Topological quantum 
computing makes use of the Aharonov-Bohm effect
in which a particle that travels around a solenoid acquires 
a phase that depends only on how many times it has encircled 
the solenoid. This topological property is highly insensitive
to disturbances in the particle's path, which leads to the 
intrinsic robustness of topological quantum computing.  
Universal topological quantum computation requires non-abelian 
Aharonov-Bohm effects, but few have been found in nature, and 
all of these are unsuitable for quantum computation. Researchers 
are working to engineer such effects, but even the most basic 
building blocks of topological quantum computation have yet to 
be realized experimentally in the laboratory.
In the long term, the robustness properties of topological quantum
computing may enable it to win out over other approaches. In the
meantime, it has inspired novel quantum algorithms. 

\subsection{What if quantum mechanics is not quite correct?}

Physicists do not understand how to 
reconcile quantum mechanics with general relativity. A complete physical
theory would require modifications to general relativity, 
quantum mechanics, or both. Modifications to quantum mechanics
would have to be subtle; the predictions of quantum mechanics 
hold to great accuracy.
Most predictions of quantum mechanics will continue to  hold, 
at least approximately, once a more complete theory is found. Since no
one knows how to reconcile the two theories, no one knows what, if any,
modifications would be necessary, or whether they would affect the
feasibility or the power of quantum computation. 

Once the new 
physical theory is known, its computational power can be analyzed. 
In the meantime, theorists have
looked at what computational power would be possible if certain changes
in quantum mechanics were made. So far these changes imply greater
computational power rather than less.
Abrams and  Lloyd showed that if quantum mechanics were 
non-linear, even slightly, 
all problems in the class $\#P$, a class that contains all NP problems
and more, would be solvable in polynomial time. Aaronson showed that 
any change to one of the exponents in the axioms of quantum mechanics 
would yield polynomial time solutions to 
all PP problems, another class containing NP. 
With these results 
in mind, Aaronson suggests that limits on computational power should 
be considered a fundamental principle guiding physical theories, much
like the laws of thermodynamics. 

\section{Conclusions}

Will scalable quantum computers ever be built?
Yes.  Will quantum computers eventually replace desktop computers?
No.  Quantum computers will always be harder to build and maintain than 
classical computers, so they will not be used for the many tasks that 
classical computers do equally efficiently.
Quantum computers will be useful for a number of specialized tasks.
The extent of these tasks is still being explored.

However long it takes to build a scalable quantum computer and whatever
the breadth of applications turns out to be, quantum information processing 
has changed forever the way in which quantum physics is taught and 
understood. 
The quantum information processing view of quantum mechanics 
clarifies key aspects of quantum mechanics such as
quantum measurement and entangled 
states. The practical consequences of this increased 
understanding of nature are hard to predict, but 
they can hardly fail to profoundly affect technological 
and intellectual developments in the coming decades.

\section{Glossary}

\hspace{1in}

{\bf Authentication} protocols are cryptographic protocols used to
establish that some or all of the commmunicating parties are who
the other parties believe them to be.

{\bf Entanglement} is a property of quantum states that does not exist
classically. Two or more subsystems of a quantum system are said to
be {\bf entangled} if the state of the entire system cannot be 
described in terms of a state for each of the subsystems. For entangled
states, the state of the subsystem is not well-defined. EPR pairs and
Bell states are the most well-known entangled states.

The {\bf no cloning principle} of quantum mechanics states that it
is not possible to create a device that reliable copies unknown
quantum states.

An algorithm is {\bf polynomial-time} in the input $n$ if the amount of
resources it uses is no more than a fixed polynomial of $n$. 

{\bf Public key encryption} is the digital equivalent of a locked 
mailbox: anyone can put a message in, but only the person with the 
key can read the message.

A proposal for quantum computers is {\bf scalable} if the amount
of resources it requires is no more than a polynomial function
of the number of qubits.

{\bf Threshold theorems} for quantum computation show that if 
the error rate can be brought below a certain threshold, arbitrarily
long and precise quantum computations can be performed.

{\bf Quantum circuits} are abstract diagrams indicating a sequence
of quantum operations to be applied as part of a computation. Quantum
circuit diagrams should not be taken to literally; they are not 
blueprints for quantum hardware. 

{\bf Quantum gates} are abstract, mathematical representations of
basic operations which can be performed on small numbers of qubits.
Sequences of quantum gates form quantum circuits. 

{\bf Quantum communication} applies quantum information processing 
to the task of communicating classical or quantum information.
Quantum teleportation and quantum dense coding are the most famous
quantum communication protocols. The former uses entangled
states and classical communication to transfer a quantum state, while
the later uses entanglement and quantum communication to communicate
classical information.

{\bf Quantum cryptography} applies quantum information processing
techniques to cryptographic applications such as key distribution,
encryption, secret sharing, and zero knowledge proofs.
Properties of quantum information, such as the no cloning principle,
provide security guarantees not available classically. 

The field of {\bf quantum information processing} examines
the theory of quantum information and its applications. Subfields 
include quantum computing, quantum cryptography,
quantum information theory, and quantum games. 

{\bf Quantum teleportation} uses entangled states and classical 
communication to transfer arbitrary quantum states from one location
to another. The reason for ``teleportation" in the name is that the
transferred quantum state is necessarily destroyed at the source by
the time the protocol is finishes, as must happen according to the
no cloning principle. Unfortunately quantum teleportation does not
enable the sort of teleportation discussed in science fiction.

A {\bf qubit}, or {\bf quantum bit}, is the fundamental unit of
quantum information, playing the role in quantum computation that
the bit plays in classical computation. While a bit has only two
possible values, a qubit has a continuum of possible values; any unit
length vector in a two dimensional complex vector space is a possible
qubit value. Common realizations of a qubit include photon polarization, 
electron spin, and a ground state and an excited
state of an atom. 

\bibliographystyle{abbrv}
\bibliography{qcCopy}

{\catcode`\/=\active \catcode`\.=\active \catcode`\-=\active
  \catcode`\@=\active \gdef\url{\tt\catcode`\/=\active \catcode`\.=\active
  \catcode`\-=\active \catcode`\@=\active
  \def/{\discretionary{\char`\/}{}{\char`\/}}%
  \def.{\discretionary{\char`\.}{}{\char`\.}}%
  \def-{\discretionary{\char`\-}{}{\char`\-}}%
  \def@{\discretionary{\char`\@}{}{\char`\@}}}} \def\annote#1{} 
  \def\tilde{\char126}
\begin{thebibliography}{10}

\bibitem{Aaronson08}
S.~Aaronson.
\newblock The limits of quantum computers.
\newblock {\em Scientific American}, 298(3):62 -- 69, Mar. 2008.

\bibitem{Aharonov-04}
D.~Aharonov, W.~van Dam, J.~Kempe, Z.~Landau, S.~Lloyd, and O.~Regev.
\newblock Adiabatic quantum computation is equivalent to standard quantum
  computation.
\newblock {\em SIAM Journal on Computing}, 37:166, 2007.

\bibitem{Bennett:1992:QC}
C.~H. Bennett, G.~Brassard, and A.~K. Ekert.
\newblock Quantum cryptography.
\newblock {\em Scientific American}, 267(4):50, Oct. 1992.

\bibitem{Carollo05}
A.~C.~M. Carollo and V.~Vedral.
\newblock Holonomic quantum computation.
\newblock {\url arXiv:quant-ph/0504205}, 2005.

\bibitem{Collins06}
G.~P. Collins.
\newblock Computing with quantum knots.
\newblock {\em Scientific American}, 294(4):56-- 63, Apr. 2006.

\bibitem{Feynman-96}
R.~Feynman.
\newblock {\em Feynman Lectures on Computation}.
\newblock Addison-Wesley, Reading, MA, 1996.

\bibitem{GRTZ-2002}
N.~Gisin, G.~Ribordy, W.~Tittel, and H.~Zbinden.
\newblock Quantum cryptography.
\newblock {\em Reviews of Modern Physics}, 74(1):145--195, Jan. 2002.

\bibitem{FeynmanHey}
A.~J.~G. Hey.
\newblock {\em Feynman and Computation}.
\newblock Perseus Books, 1999.

\bibitem{Hirvensalo}
M.~Hirvensalo.
\newblock {\em Quantum computing}.
\newblock Springer-Verlag, 2001.

\bibitem{ardaRoadmap}
R.~Hughes and et~al.
\newblock Quantum cryptography roadmap, version 1.1.
\newblock {http://qist.lanl.gov}, July 2004.

\bibitem{Koblitz-Menezes}
N.~Koblitz and A.~Menezes.
\newblock A survey of public-key cryptosystems.
\newblock {\em SIAM Review}, 46:599--634, 2004.

\bibitem{Landsburg04}
S.~E. Landsburg.
\newblock Quantum game theory.
\newblock {\em Notices of the American Mathematical Society}, 51(4):394--399,
  2004.

\bibitem{Manin80}
Y.~I. Manin.
\newblock Computable and uncomputable.
\newblock Sovetskoye Radio, Moscow (in Russian), 1980.

\bibitem{Manin07}
Y.~I. Manin.
\newblock {\em Mathematics as Metaphor: Selected Essays of Yuri I. Manin}.
\newblock American Mathematical Society, 2007.

\bibitem{HofAC}
A.~J. Menezes, P.~C. van Oorschot, and S.~A. Vanstone.
\newblock {\em Handbook of Applied Cryptography}.
\newblock CRC Press, New York, NY, 1996.

\bibitem{vanMeter06}
R.~V. Meter and M.~Oskin.
\newblock Architectural implications of quantum computing technologies.
\newblock {\em Journal on Emerging Technologies in Computing Systems},
  2(1):31--63, 2006.

\bibitem{Mosca08}
M.~Mosca.
\newblock Quantum algorithms.
\newblock arXiv:0808.0369, 2008.

\bibitem{Nielsen-05}
M.~Nielsen.
\newblock Cluster-state quantum computation.
\newblock {arXiv:quant-ph/0504097}, 2005.

\bibitem{NCbook}
M.~Nielsen and I.~L. Chuang.
\newblock {\em Quantum {C}omputing and {Q}uantum {I}nformation}.
\newblock Cambridge Press, Cambridge, 2001.

\bibitem{OBrien08}
J.~L. O'Brien.
\newblock Optical quantum computing.
\newblock {\em Science}, 318(5856):1567-- 1570, 2008.

\bibitem{Preskill-98}
J.~Preskill.
\newblock Fault-tolerant quantum computation.
\newblock In H.-K. Lo, S.~Popescu, and T.~P. Spiller, editors, {\em
  Introduction to Quantum Computation and Information}, pages 213--269. World
  Scientific, 1998.

\bibitem{Rieffel-00}
E.~G. Rieffel and W.~Polak.
\newblock An introduction to quantum computing for non-physicists.
\newblock {\em ACM Computing Surveys}, 32(3):300 -- 335, 2000.

\bibitem{Somma03}
R.~D. Somma, G.~Ortiz, E.~Knill, and J.~Gubernatis.
\newblock {Quantum simulations of physics problems}.
\newblock In {\em Quantum Information and Computation}, volume 5105, pages
  96--103, 2003.

\bibitem{Steane-97}
A.~Steane.
\newblock Quantum computing.
\newblock {\em Reports on Progress in Physics}, 61(2):117--173, 1998.

\bibitem{QIPCStrategicReport}
P.~Zoller and et~al.
\newblock Quantum information processing and communication: Strategic report on
  current status, visions and goals for research in {E}urope.
\newblock {http://qist.ect.it/}, 2005.

\end{thebibliography}
\nocite{Rieffel-00}
\nocite{NCbook}
\nocite{Steane-97}
\nocite{Hirvensalo}
\nocite{Feynman-96}
\nocite{FeynmanHey}
\nocite{Manin80}
\nocite{Manin07}
\nocite{ardaRoadmap}
\nocite{QIPCStrategicReport}
\nocite{vanMeter06}
  \nocite{OBrien08}
\nocite{Aaronson08}
\nocite{Mosca08}
\nocite{GRTZ-2002}
\nocite{Bennett:1992:QC}
\nocite{Koblitz-Menezes, HofAC} 
\nocite{Carollo05} 
\nocite{Collins06} 
\nocite{Aharonov-04} 
\nocite{Nielsen-05} 
\nocite{Preskill-98}
\nocite{Somma03} 
\nocite{Landsburg04}

Most papers on quantum computing can be found on the ePrint ArXiv
http://xxx.lanl.gov/archive/quant-ph. Two blogs frequently contain
lively discussions of recent results in quantum computation: 
\newline
http://scienceblogs.com/pontiff/ 
\newline
http://www.scottaaronson.com/blog/ 

\end{document}